# POSERS: Steganography-Driven Molecular Tagging Using Randomized DNA Sequences

Ali Tafazoli Yazdi[1], Peter Nejjar[2*] and Lena Hochrein[1*]

[1] University of Potsdam, Institute of Biochemistry and Biology, Faculty of Science, Potsdam, 14476, Germany

[2] University of Potsdam, Institute of Mathematics, Faculty of Science, Potsdam, 14476, Germany

* Corresponding authors

ORCID:

Lena Hochrein – ORCID ID: 0000-0002-2548-4318; email: hochrein@uni-potsdam.de

Peter Nejjar – ORCID ID: 0000-0001-9663-7775; email: nejjar@uni-potsdam.de

Ali Tafazoli Yazdi – ORCID ID: 0009-0006-8982-0129; email: tafazoliyazdi@uni-potsdam.de

## Abstract

Counterfeiting poses a significant challenge across multiple industries, leading to financial losses and health risks. While DNA-based molecular tagging has emerged as a promising anti-counterfeiting strategy, existing methods rely on predefined DNA sequences, making them vulnerable to replication as sequencing and synthesis technologies advance. To address these limitations, we introduce POSERS (Position-Oriented Scattering of Elements among a Randomized Sequence), a steganographic tagging system embedded within DNA sequences. POSERS ensures copy- and forgery-proof authentication by adding restrictions within randomized DNA libraries, enhancing security against counterfeiting attempts. The POSERS design allows the complexity of the libraries to be adjusted based on the customer's needs while ensuring they withstand the ongoing improvements in DNA synthesis and sequencing technologies. We mathematically validate its security properties and experimentally demonstrate its effectiveness using Next-Generation Sequencing and an authentication test, successfully distinguishing genuine POSERS tags from counterfeit ones. Our results highlight the potential of POSERS as a long-term, adaptable solution for secure product authentication.

## Introduction

Steganography is the practice of concealing the existence of a message within seemingly ordinary data to maintain the confidentiality of the message. The origins of steganography can be traced back to 440 BC, where it was employed to hide messages on writing tablets. Today, steganography is utilized in the transmission of digital multimedia data, including images, audio, and video. The decreasing cost of DNA synthesis and sequencing has expanded the use of DNA as a medium for steganography, enabling new applications in secure information embedding and molecular data storage[1,2]. This development benefits another area that deals with the use of DNA as a means of labelling products to securely track and authenticate them. Considering that counterfeiting occurs in many different industrial sectors, such as clothing, footwear, luxury items, vehicles but also pharmaceutical industry it quickly becomes clear that this not only leads to a financial loss for the industry, but also to a health risk for the end consumer[3–5]. Thus, various approaches to DNA-based molecular tagging have been reported in recent years to increase anti-counterfeiting protection compared to traditional means such as UPC barcodes, QR codes, RFID or watermarks[6].

DNA-based molecular labelling uses DNA sequences to mark high-value or security-sensitive items visibly, and more importantly, invisibly, so that they can be identified, tracked and protected from tampering. The available methods essentially differ in the types of DNA molecules and the readout of DNA sequences used to proof authenticity of the analysed sequence. Visual validation can be realized by the hybridization of two complementary single-stranded DNA sequences. If the correct DNA sequence is present on the labelled object, it hybridizes with the added reporter, resulting in specific fluorescence. In 2021, Berk *et al.* described such a visual authentication system based on a toehold-mediated DNA strand displacement[7]. Here, a fluorescent signal occurs when a quencher-modified oligonucleotide, which is not fully complementary to the fluorophore-labelled reporter, is replaced by a fully complementary oligonucleotide strand, the taggant. The spatial separation of the quencher and the fluorophore results in a fluorescent signal that can be validated within seconds to minutes using a smartphone or by eye.

Other methods use DNA sequencing to distinguish between counterfeit items without correct DNA and genuine products with the original DNA. Doroschak et al. presented their molecular tagging system Purcupine in 2020, which is based on short synthetic DNA strands, so called molecular bits (molbits)[8]. These molbits can be identified via nanopore sequencing in around 1-3 minutes, as each sequence generates unique structures that can be analysed from raw nanopore signal without the need for basecalling.

Chemical unclonable functions (CUFs) based on operable random DNA pools were introduced by Luescher et al. in 2024[9]. These CUFs utilize pools of up to $10^{10}$ unique sequences. When challenged with specific PCR primers as input, the CUF produces a distinct set of random sequences from the pool as an output response. This output can be identified using either next-generation sequencing (NGS)[9] or, in a simplified workflow, Sanger sequencing[10] with electropherogram comparisons.

Current DNA-based molecular tagging approaches utilize defined sets of DNA sequences, ranging from single sequences to over $10^{10}$ DNA sequences per product. These methods typically apply the same set of sequences to all products within a batch. Their security relies on the current limitations of DNA synthesis and sequencing technologies, leading to claims of being forgery-proof due to difficulties in amplification or direct sequencing. However, if a forger is willing to accept the increased effort, they can use advanced molecular biology techniques to identify and replicate DNA tags using methods that are already available today. Rapid advancements in DNA sequencing and synthesis techniques may compromise the long-term security of these methods. Given these limitations, existing DNA tagging methods may not offer complete protection against sophisticated counterfeiting attempts, especially not in the future. Here, we propose a steganographic method using **p**osition-**o**riented **s**cattering of **e**lements among a **r**andomized **s**equence (POSERS) to overcome these limitations of DNA-based molecular tagging systems.

## Results

## DNA protection: the secure design of POSERS

The POSERS system presents a steganographic approach to designing and producing DNA-based molecular tags, utilizing DNA libraries that contain a large number (typically millions) of unique sequences, each embedding a specific forgery-proof design. This design prevents a forger from deciphering or copying the sequence, while still allowing the designer to reliably authenticate the DNA library. In simple terms, it works by selectively excluding certain combinations of nucleotides within an otherwise randomized DNA sequence.

Depending on the respective security requirements of the customer (e.g. product lifespan, batch size, technological advancements), we define the characteristics of the individual POSERS design, which will be applied on one product batch. Remarkably, each product of one batch is labelled with a unique set of sequences and the DNA-libraries on the products differ from each other, even though following the same POSERS design. To define a design, we first determine a fixed length ($L$) of the DNA sequences, which represents the number of nucleotides in each DNA strand within the design. Next, we define a number of restricted positions ($K$), each containing only one, two or three defined nucleotides (Figure 1A). All other positions are randomly filled with an equal distribution of all four nucleotides. The selection process for the $K$ restricted positions follows a structured approach: From the available $L$ positions, $K_1$ positions are chosen, where only one nucleotide is allowed. Next, from the remaining $L - K_1$ positions, $K_2$ positions are selected, for which two bases are allowed. Finally, among the remaining $L - K_1 - K_2$ positions, we pick $K_3$ positions and for each picked position three bases are allowed. All other unselected positions will contain all four bases. For all restricted positions, both their locations and their type of restriction ($K_1$, $K_2$ or $K_3$) remain fully undisclosed and only available to the designer.

Considering the current state of DNA synthesis, we propose oligo pool synthesis for generating a POSERS DNA library[11] (Figure 1B). In this approach, each restricted position in the design ($K_1$, $K_2$ or $K_3$) will correspond to the synthesis of a single position oligo library (SPOL), where either one, two or three defined nucleotides are allowed at a certain position. In total, we have $K$ SPOLs, which is the sum of $K_1$-, $K_2$- and $K_3$SPOLs. These SPOLs are mixed to create the combined positions oligo library (CPOL).

To illustrate this approach, we selected the following experimental example for a POSERS design: The design incorporates single-stranded DNA sequences of a length ($L$) of 40 nucleotides with ten restricted positions $K_1$ and ten restricted positions $K_2$. This results in 20 SPOLs, that are combined to create the final CPOL, concealing the restrictions from individual positions. This ensures that analysing the distribution of all four nucleotides at each position in the CPOL does not reveal the originally restricted positions. A CPOL can be used either to tag a single product (e.g., a high-value product with a long lifespan) or an entire batch of products.

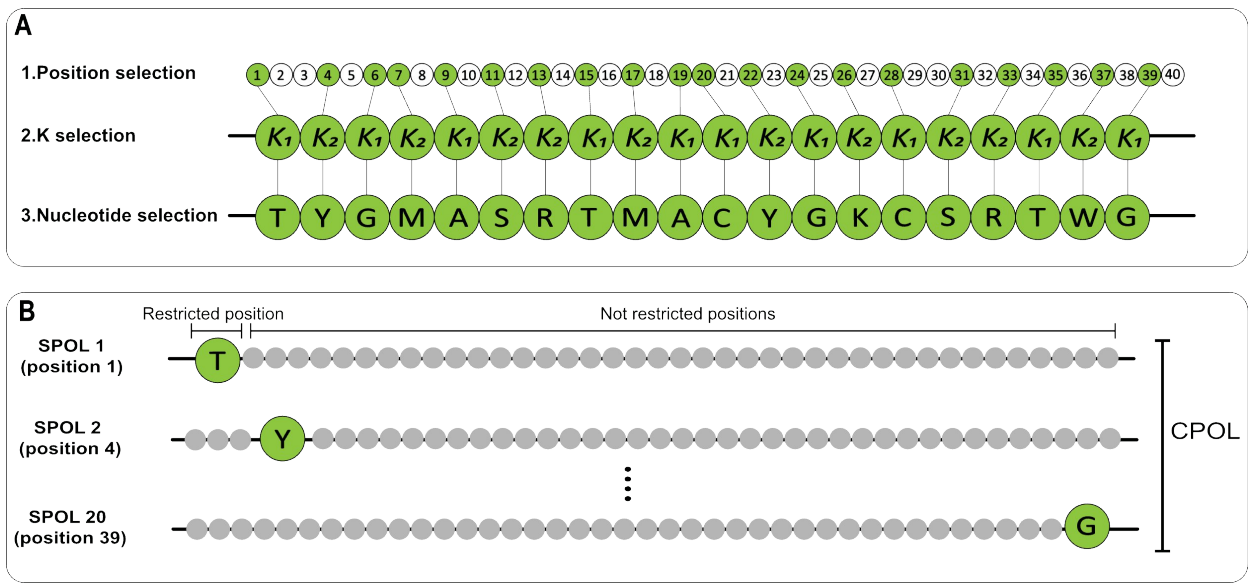

**Figure 1: POSERS design parameters and the suggested SPOL and CPOL synthesis. (A)** Diagram of generating a POSERS design. First row: Each DNA strand has a length of 40 nucleotides (illustrated as circles) with 20 restricted positions (green circles). Second row: The 20 restricted positions are defined to hold one nucleotide ($K_1$) or two nucleotides ($K_2$). Third row: For each $K_1$ position a single nucleotide and for each $K_2$ position two nucleotides are selected (Y: C or T; M: A or C; S: C or G; R: A or G; K: T or G; W: A or T). **(B)** Diagram of oligo pool synthesis of the POSERS library. For each restricted position in the design, a unique SPOL library is synthesized. In each SPOL, the picked position (green circle) only contains the allowed nucleotides, while all other positions (grey circles) can contain all four nucleotides. All 20 SPOLs generated in the pool synthesis are combined to create one CPOL.

## Exposing counterfeits: Distinguishing authentic and non-authentic libraries

For the authentication of a certain product, which is tagged by a POSERS DNA library, DNA needs to be extracted from the product and sequenced via NGS. It is evident that if a DNA sample does not meet the general criteria of the POSERS design—such as the length of the DNA sequences, and the positions and exact sequences of the constitutive sequences used in the library— it is immediately classified as non-POSERS and rejected. However, if the sample exhibits the general characteristics of a POSERS design, we proceed to assess the authenticity of a tagged product by further analysing the sequences of the CPOL.

Firstly, we perform the sample combination (SC) test: a DNA sequence is considered as authentic, if it either belongs to one of the $K$ SPOLs, or does not include any restricted combination. Whereas we will consider a sequence as not authentic, if it either does not belong to any of the $K$ SPOLs or includes any restricted combination. The SC test relies on finding restricted combinations in a non-authentic tag.

To determine the SC test sample size, we need to first calculate the probability of encountering a restricted combination in a non-POSERS DNA library. To do so, we use the example of a randomized library as a non-authentic tag as it might represent the most probable scenario for a forged DNA tag. Since a POSERS design excludes certain combinations of nucleotides at restricted positions, the resulting DNA library is smaller compared to a fully randomized library without any restrictions. As a result, non-authentic sequences will be included if a forger adds a randomized library instead of a POSERS library. A library is considered a forgery if the analysis yields a sufficiently high number of non-authentic sequences. To determine the number of sequences that must be analysed to reliably detect a forgery, we computed the absolute numbers and the proportion of non-authentic and authentic sequences among all possible sequences that would result from a randomized library. The theoretical proportion of non-authentic sequences, the missing rate $p$ equals:

$$(1) \quad p = \left(\frac{3}{4}\right)^{K_1} \left(\frac{1}{2}\right)^{K_2} \left(\frac{1}{4}\right)^{K_3}.$$

Since any sequence is either authentic or non-authentic, the proportion of authentic sequences equals $1 - p$. Thus, among the $4^L$ possible sequences, there will be $p4^L$ non-authentic sequences, and $(1 - p) 4^L$ authentic sequences. For the exemplary POSERS design in this study, the missing rate $p$ equals $5.4994 * 10^{-5}$. Therefore, we calculate the number $n$ of DNA sequences that need to be analysed in a randomized library to ensure, with very high probability, the detection of a non-authentic sequence.

$$(2) \quad \mathbf{n} = \frac{\ln(\varepsilon)}{\ln(1-\mathbf{p})}.$$

For the current CPOL, we choose the parameter ε, which represents the possibility to not detect a fully randomized library as a forgery, to be 0.00001%. In this way, the number of sequences equals $n = 2.5121 * 10^5$. Thus, we define $n$ as the number of sequences the designer must test to ensure that the library can be reliably authenticated and distinguished from a randomly generated library. The mathematical verification of equations (1) and (2) is provided in the Supplementary Information.

DNA synthesis and sequencing may introduce errors and biases that deviate from the ideal mathematical assumptions[12,13]. To ensure the secure authentication of a POSERS library, the errors of the experimental steps must be taken into account (Figure 2). For this purpose, an exemplary CPOL was synthesized by the company Integrated DNA Technologies (IDT) using oligo pool synthesis service. The transposase used for sequencing library preparation in this study has previously shown to remove approximately 50 bp from each end of a DNA sequence[14]. To ensure a sufficient length for sequencing of the CPOL library, DNA sequences contain 40 nucleotides of the designed sequences flanked by 80 nucleotides of fixed constitutive sequences on both the 5' and 3' ends. As a negative control, the same library is ordered, but with the design replaced by 40 nucleotides of randomized sequence. The negative control is also generated using a pooled synthesis approach, alongside a second pool sample in which the design is replaced by a synthetic constitutive sequence. The constitutive DNA sample was used solely as proof of successful sequencing sample preparation and was not included in further analysis. Both libraries are ordered as single-stranded and are converted to double-stranded libraries prior to testing. For sequencing, we used the Illumina sequencing platform 2*150 with a PCR-free kit to minimize the effect of PCR duplication on the original sample. The sequencing was performed as a paired-end run. However, since a single read covers the full length of the design, the analysis is based on single reads. Twenty-five nanograms of both the double-stranded CPOL and the double-stranded control sample were used for sequencing, as this is the minimum sample quantity recommended by the PCR-free kit protocol.

The raw data obtained from the sequencing experiment was analysed by a custom program (see Methods) to filter out the duplicated reads and reads that do not follow the CPOL design with respect to the length and correct primer binding sites. This resulted in 1,029,652 unique sequences from the CPOL sample and 468,156 sequences from the random control sample. The POSERS authentication program subsequently finds all the restricted combinations in these sequences according to the design (see Methods). As expected, the program did not detect any forbidden combinations in the original sample. In contrast, 29 forbidden combinations were identified in the randomized control sample (Supplementary Table 1). This means that the probability of a sequence being forbidden in a library of completely random sequences can be estimated as 29 divided by 468,156, which equals $6,194516 * 10^{-5}$. Since the estimated probability is close to, and even slightly higher than the mathematically calculated probability ($p= 5.4994 * 10^{-5}$), it can be concluded that experimental errors from DNA synthesis and sequencing do not negatively impact the effectiveness of the SC test in identifying restricted combinations. Furthermore, the original CPOL could be securely distinguished from the randomized control sample.

Secondly, to ensure that the sequencing result of the test sample accurately represents the complexity of the original sample, we recommend conducting the sample variety (SV) test performed by the POSERS authentication program. Here, we consider a sample as authentic if it follows the expected design variety and includes all allowed nucleotides in each SPOL. By isolating sequences from the sequencing data that can only result from a single SPOL, we can analyse whether all allowed nucleotides are present at the restricted position. This is achieved by selecting sequences in which all positions, except for the position of interest, contain a restricted nucleotide. This ensures that each restricted position includes all allowed nucleotides and reveals a scenario in which a forger correctly predicts the position of a restriction but not the type of restriction ($K_1$ instead of $K_2$ or $K_3$). For example, a sequence can be clearly assigned to SPOL 22 if all other positions contain nucleotides which are excluded in their respective SPOL. By examining all positions in the CPOL sample, we confirmed that all allowed nucleotides are present at the restricted positions where SPOL specific sequences are found (Supplementary Table 2). This test therefore serves as proof for the diversity of the sample. Thus, a DNA tag imbedded within a POSERS design can be reliably authenticated by a combination of the SC and SV tests.

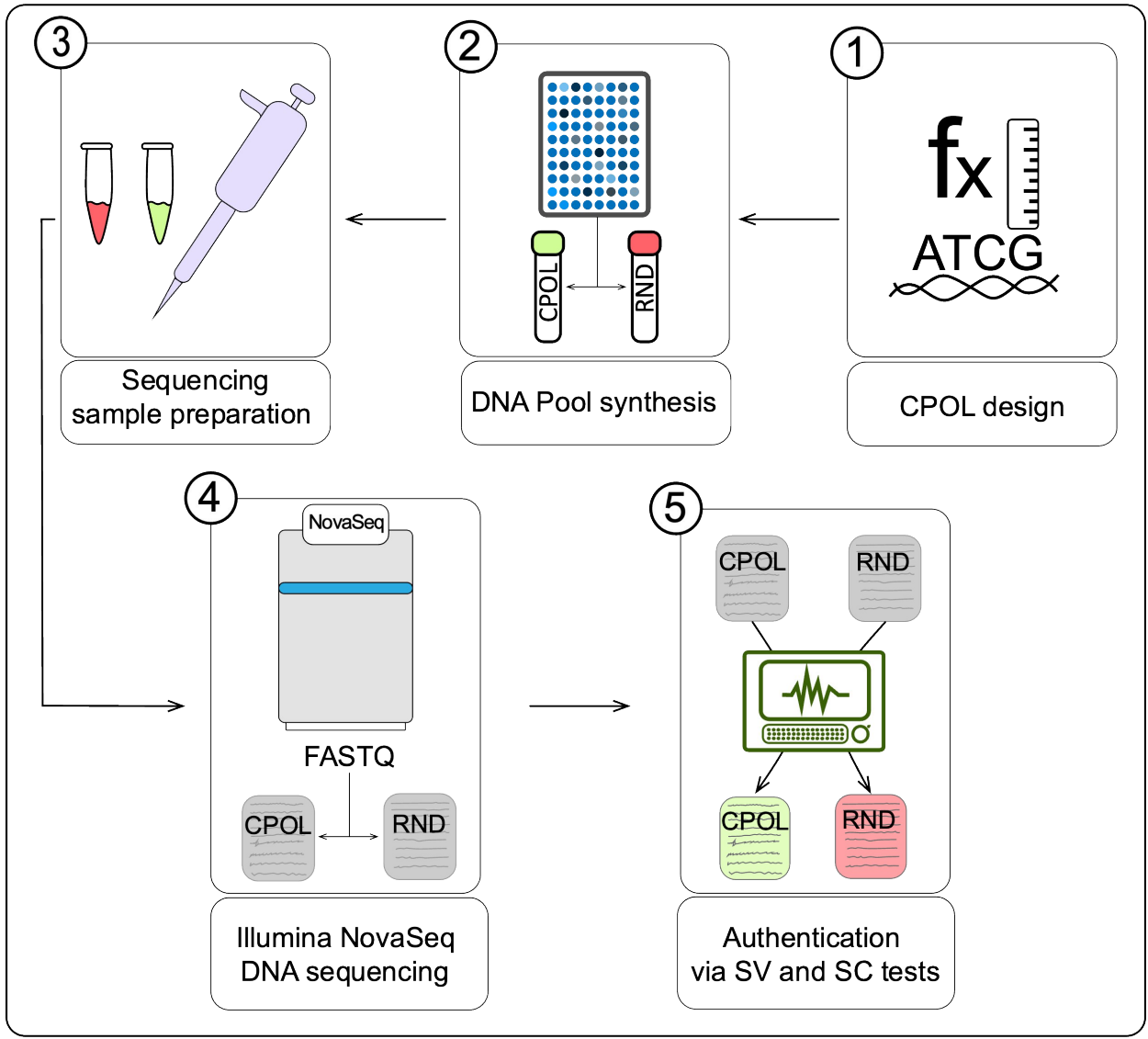

Figure 2: Schematic overview of the experimental design for generating and authenticating a POSERS DNA library. 1) A POSERS design is selected considering factors including the length of DNA sequences (L), the number of restricted positions ($K$), and the nucleotides allowed in each of the $K$ positions. 2) The CPOL sample and a control (fully random library) are synthesized through separate DNA pool syntheses. 3) Samples are prepared for Illumina sequencing by double-stranding and library preparation using Illumina DNA PCR-Free prep. 4) The sequencing is performed on the Illumina NovaSeq platform, generating FASTQ results. 5) The FASTQ results from the CPOL and random samples are checked using the authentication program. After performing the CS and SV tests, the CPOL sample is approved, while the random sample is rejected.

## Breaking the design: How forgers might attack the library

If forgers attempt to counterfeit products tagged by POSERS, they can pursue two strategies: either multiplying the existing DNA library or identifying the design of the POSERS tags to synthesize a forged DNA library.

## First angle of attack: Copying the sample

The first approach a forger could take to generate authentic DNA sequences is to gain access to a number of authentic products, denoted as $R$. As each product is tagged with at least $n$ sequences, the forger has access to $R*n$ authentic sequences, which they may proceed to use for copying the authentic sequences. Given the current state of DNA synthesis and sequencing, there are two options available for copying the DNA sample:

(1) Synthesizing a new copy of isolated and identified sequences

The first approach a forger could take is to isolate DNA from the accessible original product and perform a sequencing experiment to identify as many authentic sequences as possible. Subsequently, the sequencing data can be used to resynthesize the authentic sequences. However, the built-in variety of the design guarantees that synthesizing such a large number of sequences (at the moment >100,000 unique sequences per product) individually remains entirely impractical. While the cost of DNA synthesis decreases overtime, the POSERS design allows for an increase in sequence variety as needed, ensuring that this approach remains unfeasible. Remarkably, if the same library of synthesized fragments will be applied on several products, the authentication test will identify the forgery by tracking duplicate sequences.

(2) Duplicating an original POSERS library by amplification

The second approach to copying a POSERS library involves amplifying DNA isolated from the original product using PCR. To prevent this, we implemented technical hurdles into the POSERS design that restrict high-efficiency PCR amplification of the POSERS library: First, the POSERS library will be designed as single-stranded DNA sequences with only a single primer binding site. This would allow the designer who knows the primer binding site to convert the single-stranded library efficiently into a double-stranded DNA library but prevents the forger from directly amplifying the library using PCR. Additionally, providing the sample as single-stranded DNA further complicates the identification of the primer binding sequence which is necessary to generate a double-stranded library. To bypass this restriction and amplify the original library, a forger would need to perform multiple enzymatic steps: attaching a single-stranded oligo tag carrying another primer binding site to the other end of the single-stranded POSERS sequence[15], PCR amplify or subclone the resulting DNA fragments and finally remove the added oligo tags from all DNA sequences without leaving any detectable scar sequences. This process is not only highly resource-intensive, but the multiple enzymatic reactions required for the attachment and subsequent removal of primer binding sites are also likely to leave detectable residues at the ends of the POSERS design and operate with very low efficiency. As a result, the forged library exhibits low diversity and contains unwanted sequence residues that can be detected during authentication. This makes it clear that even if a forger would take the effort and still copy the POSERS library by PCR, the authentication program will identify this copy due to a lower variety of sequences and detectable residues from enzymatic reactions.

We initially assumed that the presence of duplicated sequences could serve as an additional measure for authenticating a POSERS tag, given that a forged PCR-amplified sample would exhibit a higher number of duplicated sequences due to its lower diversity. Since only a fraction of the theoretically possible sequences from the total POSERS design is synthesized, we expected an authentic POSERS tag to be free of duplicates. This led us to consider duplication as a potential indicator of forgery. However, previous studies have reported a high rate of false duplicate sequences, known as optical duplicates, as an artifact of Illumina sequencing[16]. This poses a challenge for authenticators in distinguishing PCR-induced duplicates from sequencing-related duplicates when using the NGS approach applied in this study.

Consistent with these findings, our raw sequencing results of both the CPOL and random control revealed that over 20% of the reads were duplicated, with the majority of these duplicates occurring only once (Figure 3). This effect prevented us from detecting duplicates caused by PCR amplification of our original CPOL sample (see Methods). Additionally, sequencing results from the PCR-amplified sample exhibited a similar pattern of duplicated reads, even after filtering optical duplicates using the Clumpify (bbmap) tool (https://sourceforge.net/projects/bbmap).

However, selecting alternative Illumina sequencers or exploring alternative NGS platforms such as Nanopore and PacBio could enable the identification of duplicates as an additional measure for detecting a copied library, along with testing for the lower diversity and sequence residues.

Remarkably, the challenge of detecting duplicates arises only when the forged PCR-amplified library is applied to a single product. If the forger uses the PCR-amplified tag across multiple products, these products are expected to undergo independent sequencing runs. In this scenario, no optical duplicates should be present between runs. Therefore, any duplicated sequences identified across different sequencing runs can be attributed to PCR amplification, indicating that the tag was forged. Consequently, all products sharing the duplicated sequences can be classified as counterfeit.

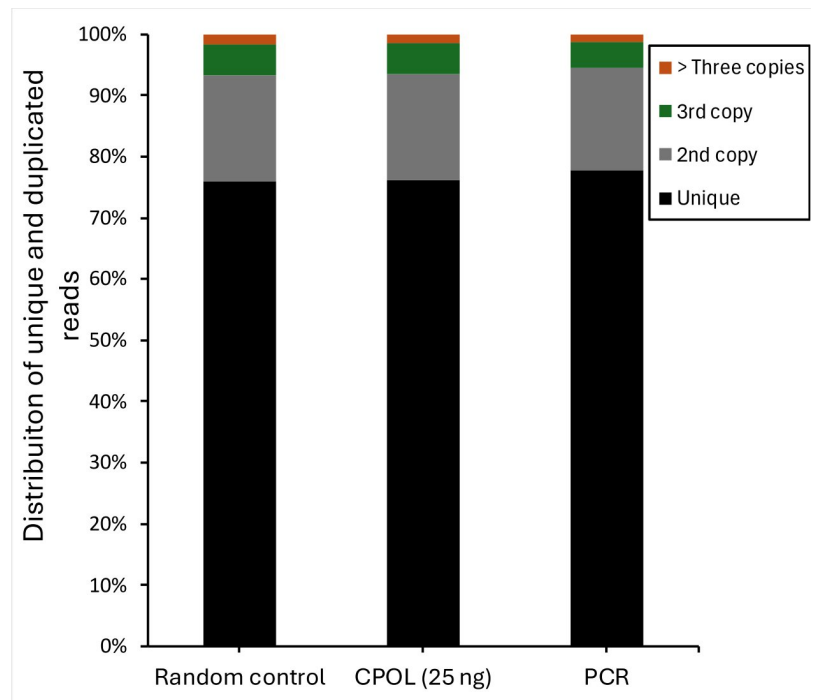

**Figure 3: Duplication report in the FASTQ sequencing results**. All three library samples exhibit a similar duplication pattern. About 75% of the reads in the raw FASTQ file are unique. Approximately 17% represent the second occurrence of a unique read, while the third copy accounts for about 5% of all reads. The remaining reads correspond to the fourth and subsequent copies of a unique read in the dataset.

## Second angle of attack: Finding the design

The second angle of attack is to find the specific design, meaning the number and kind of restricted positions, by sequencing an authentic DNA library. This approach would be the only way for the forger to synthesize a POSERS library in a cost-effective manner. To do so a forger has two general options:

(1) Analysing distribution differences of nucleotides at individual positions

Forgers could try to detect irregularities in the nucleotide distribution at certain positions of the DNA sequence caused by the implementation of the POSERS design. The implementation of a POSERS design in a nucleotide library results in a slight bias in nucleotide distribution at the picked restricted positions, while positions without restrictions have no bias with an expected theoretical proportion of 25% for each of the four nucleotides. However, at a position which has been picked, the expected proportion of an allowed letter is different. Namely, at a position where $i = 1,2,3$ letters are allowed in the design, the expected theoretical proportion of an allowed letter is

$$(3) \qquad \frac{1-1/K}{4} + \frac{1}{iK}$$

whereas the not-allowed letters have an expected proportion of

$$(4) \qquad \frac{1-1/K}{4}$$

The mathematical verification of equations (3) and (4) is provided in the Supplementary Information. Applied to our POSERS design with 20 restricted positions and two allowed nucleotides, a deviation of 1.25% from the average results in a proportion of 26.25% for the allowed nucleotides and 23.75% for the not-allowed nucleotides. For a scenario with one allowed nucleotide and 20 restricted positions, a deviation of 3,75% leads to a proportion of 28.75% for the allowed nucleotides and 23.75% for the not-allowed nucleotides.

Based on the given mathematical assumption, we proceed with predicting how a forger might attempt to reconstruct the CPOL design. Assuming the forger was able to successfully sequence the original CPOL from an authentic product and attempts to predict a design based on the assumption that the design has 20 restricted positions:

We calculate the average nucleotide distribution across all 40 positions in all reads based on the FASTQ data, resulting in the following distribution: A: 20.68%, T: 29.57%, C: 18.83%, and G: 30.93%. We designate positions where the nucleotide distribution exceeds the average by 1.25% (deviation from average calculated for $K$ = 20, $i$ = 2) as $K$ restricted positions (Supplementary Table 3). If a forger attempted to predict the design using the suggested approach, their prediction would result in the pattern shown in Figure 4A. However, the forger would only be able to identify 17 restricted positions correctly, leading to nine incorrect predictions in the forged library. Before investigating how false predictions can be identified in the forged library, we first categorize the two different types of falsely predicted positions (Figure 4B):

(a) The first type of false prediction introduces incorrect nucleotide combinations into the sample. This occurs when a position without restrictions is falsely predicted as a restricted position (FNP). In FNP scenario, an additional SPOL will be generated by the forger introducing a restriction at a false position, while the other positions have no restriction. This SPOL introduces all false combinations that a fully randomised library introduces. Another possible false prediction occurs when a position with restriction is not identified or identified with at least one nucleotide restricted from this position (FPN). The predicted design includes three FPNs (Figure 4B). In FPN scenario, the forger either excludes a SPOL from the CPOL, resulting in a random distribution of all four nucleotides at this position, or includes a SPOL with at least one wrong nucleotide included. Consequently, this position contains nucleotides that are excluded by the design, leading to restricted combinations in the forged sample that do not exist in the authentic sample.

The introduced restricted nucleotide combinations resulting from FNP and FPN can be identified by the SC test. However, the non-authentic library with a certain number of correctly predicted positions incorporates fewer false combinations in comparison to the randomized library. This should be considered to adjust and increase the number of sequences $n$ that needs to be analysed, calculated from equation (2). This means, as soon as a single FNP or FPN SPOL is present in a CPOL, we can compute the probability that a sequence from the CPOL is not authentic as $p' \geq 2 * \frac{p}{3K}$. This means that the number of sequences we need to test from a POSERS library is calculated by multiplying the number $n$ from equation (2) by $\frac{3K}{2}$. For the suggested CPOL design, this means that the number of sequences to be tested increases from $2.5121 * 10^5$ sequences per library to $7.5363 * 10^6$ to ensure the detection of an FNP or FPN in a forged sample.

The downside of increasing the number of $n$ is the higher cost associated with incorporating more DNA into the tag and the increased sequencing requirements. However, if necessary, adjusting $n$ allows for maximizing the security level of authentication for the POSERS tag, ensuring that even false combinations introduced by a single FPN/FNP prediction can be detected by the SC test.

(b) The second type of false prediction occurs when the forger correctly identifies the restricted position but fails to determine all allowed nucleotides at that position (FHP). For example, if a designed position allows two nucleotides, but the forger incorrectly restricts one of them, no false combinations are introduced into the library. However, all DNA sequences generated from the forged SPOL will lack the missing nucleotides at that position. As a result, the FHP can be detected by our authentication program using the SV test. In the predicted design, six FHP predictions were identified (Figure 4A).

Therefore, unless the forger accurately predicts all restricted positions and all allowed nucleotides at these positions, the forged sample will either contain false predictions or limited diversity, which can be traced back by the POSERS authentication tests and used to distinguish it from the original sample. Furthermore, the variables during DNA synthesis, such as the composition of the nucleotide mixture and the techniques used for wobble oligo synthesis and DNA pooling synthesis, can vary from sample to sample, as they differ between synthesis companies.

Therefore, a forger would need access to the specific DNA synthesis information for each sample of one batch to develop an optimal prediction strategy. Without detailed knowledge of the POSERS design and the exact synthesis method, it would be extremely difficult for a forger to develop a more precise or effective prediction method than the one demonstrated in this study. Furthermore, the POSERS designer can make prediction even more challenging by adjusting the ratio of individual SPOLs within the design. In the CPOL design presented here, it is assumed that all SPOLs are present in equal proportions within the sample. However, by modifying the DNA pooling after synthesis, the designer can alter the relative representation of individual SPOLs in the final CPOL, further complicating the prediction process. Finally, the designer can fine-tune the design parameters, such as increasing the number of restricted positions. This, in turn, raises the number of SPOLS, thereby minimizing deviation from the average calculated using equation (3) and making prediction significantly more difficult.

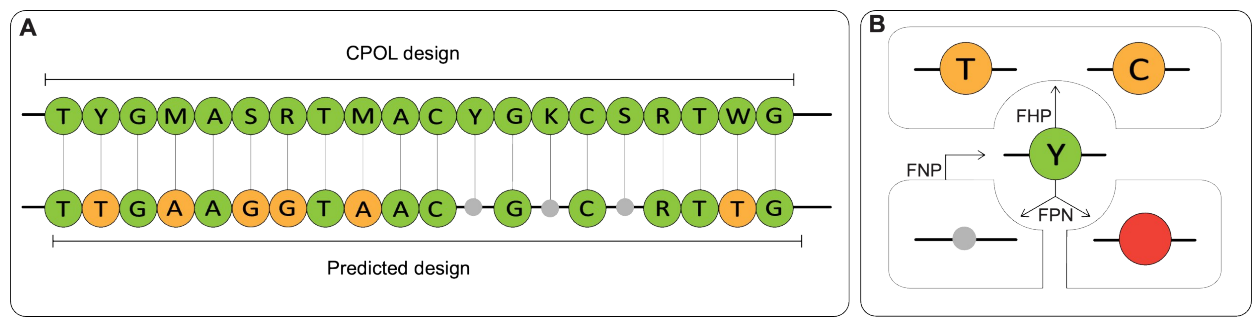

**Figure 4: False design prediction from nucleotides distribution analysis**. (**A**) The predicted design based on the nucleotide distribution analysis includes three FPN predictions (grey circles) and six FHP predictions (yellow circles). (**B**) FPN prediction: a restricted position (here Y = C or T, depicted as green circle) is not recognized (small grey circle) or the prediction for this position includes a forbidden nucleotide (red circle). FNP prediction: a position with no restriction in the original design is considered as a position with restriction. FHP: not all nucleotides allowed for a certain restricted position are recognized.

(2) Check the combinations of restricted positions

The second strategy for deciphering the design involves analysing all nucleotide combinations in the sample to identify abnormal restrictions, such as the absence of certain nucleotide combinations. When analysing the nucleotide combinations, the key difference between the designer and the forger lies in the number of sequences that need to be examined. The designer knows the exact locations of restricted positions and which restrictions to check, significantly reducing the number of sequences that need to be analysed. This number, defined as the lower threshold, is the minimum number of sequences that need to be applied on one product ($n$ calculated from equation (2)). In contrast, the forger lacks this information and must analyse a much larger number of sequences to identify the restrictions. This upper threshold is set by the number of sequences a forger would have to analyse to identify the POSERS design. Applying fewer sequences as these number per product ensures that the design remains secure and cannot be deciphered. While in a real scenario the forger would not have knowledge of all restricted combinations, we assume here that future methods could potentially generate this information. In that case, the forger could analyse the number of produced combinations only at the $K$ picked restricted positions, e.g. by systematically going through all values for $K$ and all the $\binom{L}{K}$ possible ways to pick $K$ positions. In this way, when K reaches 20, the forger might realize that $p4^K$ sequences are never produced and conclude that these combinations are forbidden. Since $K < L$, the forger needs access to fewer sequences to make this conclusion. Therefore, the number of DNA sequences at the upper threshold must be limited so that for the $K$ restricted positions, no more than $(1-p)4^K$ combinations are possible. The simplest way to achieve this is to produce fewer than $(1-p)4^K$ DNA sequences per design and in this way limit the amount of available DNA sequences, which will prevent a counterfeiter from finding the design. While we use the simplest approach here to define the upper threshold, mathematical justification supports the generation of even more sequences while still ensuring the safety of the design (Supplementary Information).

The security resistance of the POSERS design against the two forgery attempts presented here demonstrates its effectiveness in generating a forgery-proof and copy-resistant DNA tag.

## Defining the Application: How many products can be tagged with one POSERS design

The defined upper and lower threshold is used to calculate the number of unique products that can be tagged with a single POSERS library following one design. For the CPOL design defined here, the lower threshold is calculated to be $n = 2.5121 * 10^5$. The number of products $P$ is then constrained by the inequality $Pn \leq (1-p)4^K$, such that

$$(5) \quad P \leq \frac{(1-p)\,4^K}{n}.$$

In this way, the minimum number of products that can be tagged by the current CPOL is $P = 4.3766 * 10^6$. This demonstrates the potential of a single POSERS design to tag a large number of products before a design change becomes necessary. In addition to the mathematical calculations, these findings must be experimentally validated to assess their practical feasibility. This includes determining the minimum amount of DNA required for sequencing to reach the lower threshold of reads. Additionally, the impact of applying DNA to a product—a paper

substrate—as well as the subsequent extraction process on the sequencing results needs to be examined. Therefore, we prepared seven dilutions of the double-stranded library, ranging from 0.01 ng to 25 ng. All samples were multiplexed and used in one sequencing run (see Methods section). The results indicate that using 5 ng of the CPOL can yield sufficient sequencing results to cover the $2.5121 * 10^5$ reads required for the authentication process (Figure 5A). The required amount of the POSERS tag is comparable to competing DNA tagging technologies[7,17] and considering the continues decrease in DNA synthesis cost[18] can be considered commercially viable. This result is relative to the scale of the sequencing run and heavily dependent on the sample preparation method. Consequently, different methods may produce varying results, leading to differences in the amount of DNA required for authentication and tagging.

To investigate the effect of applying a POSERS library to a product, the DNA was applied to paper, extracted, and subsequently sequenced. Two samples were prepared for this test. In both cases, 25 ng of the DNA library was applied to filter paper. After drying for 24 hours, the DNA was extracted by applying water to the paper and collecting the eluate. For one sample, the extracted DNA was directly used for sequencing preparation. For the other sample, the DNA was purified using a purification kit before proceeding with sequencing preparation. The results showed that, unlike the non-purified sample, the purified sample achieved a number of sequencing reads comparable to the untreated sample that was not applied to paper (Figure 5B). Interestingly, the purified DNA extracted from the paper yielded more than twice the sequencing output compared to 25 ng of the pure DNA library. However, this difference can be explained considering errors in sample preparation handling. These findings indicate that applying DNA to paper and subsequently extracting dried DNA does not considerably impact sequencing output, provided that the sample undergoes purification after isolation.

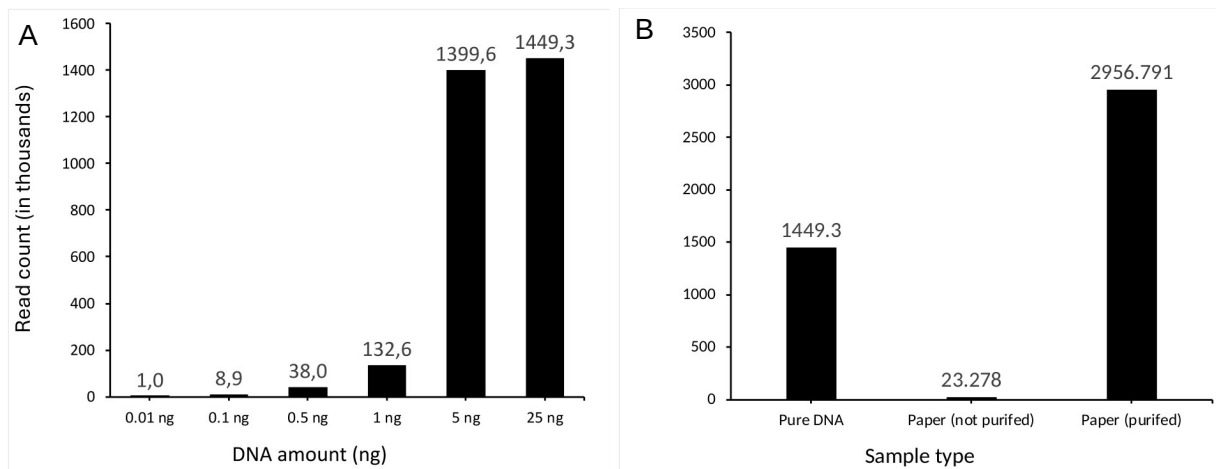

**Figure 5: Experimental impact on DNA sequencing output**. (**A**) The effect of DNA input on sequencing results, showing increasing sequencing output for CPOL samples with input amounts ranging from 0.01 ng to 25 ng. (**B**) The impact of applying the CPOL library to paper and subsequent DNA purification on sequencing output, following DNA isolation from paper.

## Conclusion

In this study, we introduced POSERS, a steganographic approach for securing DNA tags with unmatched security properties, including copy-proof and forgery-proof qualities. We mathematically demonstrated that the designed sample can be uniquely distinguished from any other DNA sample and experimentally validated these findings. Additionally, our analysis of potential forgery scenarios confirmed that directly copying a POSERS library is, first of all, highly challenging and impractical, and would only allow the tagging of a single sample. Furthermore, predicting the design from nucleotide distribution or recording all possible combinations is infeasible. The adaptability of POSERS ensures a secure and future-proof solution, capable of keeping pace with advancements in DNA synthesis and sequencing technologies. We successfully demonstrated the distinctive quality of the POSERS strategy, which, unlike existing methods, assigns each product a unique DNA library that can be authenticated with as little as five nanograms of DNA per product.

Furthermore, we demonstrated the applicability of POSERS tags on paper, confirming their feasibility for secure tagging and authentication. However, POSERS tagging application is not limited to paper. Previous research has shown that short DNA sequences can be incorporated into a wide range of materials, including lactose pills, milk, ink and other substances[19–23]. These findings highlight the versatility of DNA-based tagging and open up numerous potential applications for POSERS tags across various industries. Moreover, the POSERS design principle can extend beyond DNA, applying to other materials with precisely arranged monomers, such as proteins and synthetic compounds such as the peptide nucleic acids (PNA)[24], provided they offer sufficient diversity for effective implementation.

These results position POSERS as a robust and adaptable anti-counterfeiting strategy, with potential for widespread adoption in supply chain security and product authentication.

## Methods

## SPOL and CPOL synthesis

To generate the oligonucleotide library, we designed a test library with a length of 40 nucleotides incorporating 20 positions with restrictions and 20 positions with no restriction. 20 separate SPOL samples were synthesized using the IDT Pooled Oligo Pools service and subsequently pooled to form the final CPOL (AT251). Each SPOL contained a POSERS region, single-stranded sequences of 40 nucleotides, with one restricted position, which means that this position lacks two or three nucleotides. All other positions are randomly filled with an equal distribution of all four nucleotides. To facilitate PCR amplification and Illumina library preparation, 80 bases of a synthetic constitutive sequences were added on both the 5' and 3' ends of the designed sequence, resulting in a total oligonucleotide length 200 nucleotides.

The sequence overhangs were:

For 5' end:

ATTGACCAACACTACTAACTTACATTTAACGTCATGCAATCTTCGAGAAGCAATGACAACGATGCCTTTGGTTATTTGAT

For the 3' end:

ACTGAGATAGCAATATGATAAAGATGTTATTGAACGAGTGGAATGCATAGAGACAGGAATCGTCCTTGTACTGCGTCTAA

For the control sample, we used the same IDT Pooled Oligo Pools service to generate two libraries in which the 40 nucleotides from the design were replaced by fully randomized or constitutive nucleotides, flanked by the same 80 nucleotides constitutive sequences (AT252).

The sequence of all oligonucleotides used in this study are provided in Supplementary Table 4.

## Double-stranding of CPOLs

To determine the lowest required sample amount for Illumina sequencing, both the test CPOL (AT251) and the control oligo sample (AT252) were converted to double-stranded DNA using AT17 as the complementary oligo primer.

One µl from 100 pmol/µl dilution of both AT251 and AT17 was used for double stranding the CPOL library while 5 µl from 5 pmol/µl dilution of AT252 and AT250 used for the control sample. Both reactions were performed in 20 µl volume using PrimeSTAR® Max DNA Polymerase (Takara Bio, Saint-Germain-en-Laye, France) under the following conditions in a thermal cycler: 98 °C for 15 s (denaturation), 55 °C for 10 s (annealing), 68 °C for 90 min (extension) for one cycle. The double-stranded CPOL was then purified using the Zymo DNA Clean & Concentrator Kit (Zymo Research, Freiburg, Germany).

## CPOL sample preparation and quantification for Illumina Sequencing

The purified double-stranded CPOL sample was prepared in seven different DNA amounts, ranging from 0.01 to 25 ng, with final volumes between 1.3 to 4.2 µl. Additionally, 25 ng of the double-stranded control sample was used for sequencing.

Extraction of CPOLs from filter paper:

To test the extraction efficiency and influence of paper extraction on sample quality, two paper samples were prepared. For this, 25 ng of CPOL DNA was diluted in 10 µl of ddH$_2$O and applied to two MN 615 filter papers (each 2mm*2mm) (Machery-Nagel, Düren, Germany), which were placed inside Eppendorf tubes. The samples were left to dry for 24 hours. After drying, 10 µl of ddH$_2$O was added to each paper, and after several rounds of pipetting, the recovered liquid was transferred into new tubes. One sample was further purified using the Zymo DNA Clean & Concentrator Kit, while the other sample was used directly for sequencing preparation.

PCR sample preparation:

To analyse the impact of PCR amplification of a CPOL on the sequencing results, 1 ng of the double-stranded CPOL was used as a template, along with AT17 and AT250 as primers (Supplementary Table 4). The PCR amplification was performed using KAPA HiFi HotStart ReadyMix PCR kit (Roche, Basel, Schweiz) under the following thermal cycling conditions: 98 °C for 45 s (initial denaturation); **25 cycles:** 98 °C for 15 s, 60 °C for 30 s, 72 °C for 30 s; 72 °C for 60 s (final extension).

PCR product was purified using the Zymo DNA Clean & Concentrator Kit (Zymo Research) and 25 ng of the purified product was used for sequencing.

DNA quantification:

DNA concentration for all samples were determined by DS-11 FX+ Fluorometer (DeNovix, Wilmington, DE, USA) together with the Qubit dsDNA HS Assay Kit (Invitrogen, Waltham, MA, USA).

## Library preparation and Illumina sequencing

For library preparation, we used the Illumina PCR-free Library Preparation Kit (Illumina Inc., San Diego, CA, USA) following the "Thermal Cycler, Low Input" protocol provided by Illumina. To enable sample identification, we applied Illumina UD Indexes to barcode each sample. The specific indexes used are listed in Supplementary Table 5. Following indexing, all 11 samples were pooled into a single 1,5 ml Eppendorf tube and sent for sequencing by Microsynth AG RTL Illumina 50 Mio shotgun read pairs 2*150 service. The sequencing was run on a NovaSeq sequencer using NovaSeq v1.5 chemistry. Microsynth AG demultiplexed the sequencing result of 11 mixed samples based on the UD Indexes provided by us and delivered a separate FASTQ sequence result for each 11 samples.

## Data analysis

The FASTQ files received from Microsynth AG were analysed using our custom program designed to evaluate combinations of nucleotides at the restricted positions. For each dataset, the program identified the nucleotide combinations present at these positions and reported the number of recorded combinations that did not follow the restrictions implemented in the design.

To analyse sequence duplications, the Clumpify tool from the BBmap package was used with the default settings for the NovaSeq sequencer.

## Data availability

The sequencing data for all 11 samples are available on figshare repository: https://doi.org/10.6084/m9.figshare.28504847. Any additional data will be made available upon reasonable request.

## Code availability

The code from the customer analysis program used in this paper are available on figshare repository: https://doi.org/10.6084/m9.figshare.28505531.v1.

## Conflict of interest

The authors declare competing financial interests, due to a patent application for the POSERS design, filed by the University of Potsdam with A.T.Y. and L.H. as inventors (Patent numbers EP23700850.3 and US18/729,231).

## Author contributions

A.T.Y. and L.H. developed the overall strategy. A.T.Y., L.H. and P.N. designed the experiments. A.T.Y. conducted the experiments and analysed the data. P.N. provided mathematical proof. A.T.Y. and L.H. wrote the manuscript with contributions from P.N.

## Funding

This work was supported by Potsdam Transfer, the central institution for start-ups, innovation, and the transfer of knowledge and technology at the University of Potsdam, through the Funding of Knowledge and Technology Transfer (FöWiTec) initiative.

## Acknowledgements

We thank Dr. Yannic Vargas as well as all members of the Junior Research Group TAILOR for fruitful and inspiring discussions. We thank the Department of Molecular Biology at the University of Potsdam for technical support throughout the study. The authors used OpenAI's ChatGPT (model GPT-4o) to adjust the manuscript's English.

## References

1. Cui, M. & Zhang, Y. Advancing DNA Steganography with Incorporation of Randomness. *ChemBioChem* **21**, 2503–2511 (2020).

2. Na, D. DNA steganography: Hiding undetectable secret messages within the single nucleotide polymorphisms of a genome and detecting mutation-induced errors. *Microb Cell Fact* **19**, 128 (2020).

3. Arias Burgos, C. & Wajsman, N. *Economic Impact of Counterfeiting in the Clothing, Cosmetics, and Toy Sectors in the EU*. (EUIPO, 2024).

4. Substandard and falsified medical products. https://www.who.int/news-room/fact-

sheets/detail/substandard-and-falsified-medical-products (2024).

5.  OECD/EUIPO, *Dangerous Fakes: Trade in Counterfeit Goods That Pose Health, Safety and Environmental Risks*. (OECD Publishing, Paris, 2022).

6.  Kuzdraliński, A. *et al.* Unlocking the potential of DNA-based tagging: current market solutions and expanding horizons. *Nat Commun* **14**, 6052 (2023)

7.  Berk, K. L. *et al.* Rapid visual authentication based on DNA strand displacement. *ACS Appl Mater Interfaces* **13**, 19476–19486 (2021).

8.  Doroschak, K. *et al.* Rapid and robust assembly and decoding of molecular tags with DNA-based nanopore signatures. *Nat Commun* **11**, 5454 (2020).

9.  Luescher, A. M., Gimpel, A. L., Stark, W. J., Heckel, R. & Grass, R. N. Chemical unclonable functions based on operable random DNA pools. *Nat Commun* **15**, 2955 (2024).

10. Luescher, A. M., Stark, W. J. & Grass, R. N. DNA-based chemical unclonable functions for cryptographic anticounterfeit tagging of pharmaceuticals. *ACS Nano* **18**, 30774-30785 (2024)

11. Kuiper, B. P., Prins, R. C. & Billerbeck, S. Oligo pools as an affordable source of synthetic DNA for cost-Effective library construction in protein- and metabolic pathway engineering. *ChemBioChem* **23**, e202100507 (2022).

12. Filges, S., Mouhanna, P. & Ståhlberg, A. Digital quantification of chemical oligonucleotide synthesis errors. *Clin Chem* **67**, 1384–1394 (2021).

13. Stoler, N. & Nekrutenko, A. Sequencing error profiles of Illumina sequencing instruments. *NAR Genom Bioinform* **3**, lqab019 (2021).

14. Bruinsma, S. *et al.* Bead-linked transposomes enable a normalization-free workflow for NGS library preparation. *BMC Genomics* **19**, 722 (2018).

15. Ma, Z. *et al.* Characterization of an ssDNA ligase and its application in aptamer circularization. *Anal Biochem* **685**, 115409 (2024).

16. Bainbridge, M. N. *et al.* Whole exome capture in solution with 3 Gbp of data. *Genome Biol* **11, R62 (2010)**.

17. Meiser, L. C. *et al.* DNA synthesis for true random number generation. *Nat Commun* **11**, 5869 (2020).

18. Meiser, L. C. *et al.* Synthetic DNA applications in information technology. *Nat Commun* **13, 352** (2022).

19. Altamimi, M. J. *et al.* Anti-counterfeiting DNA molecular tagging of pharmaceutical excipients: An evaluation of lactose containing tablets. *Int J Pharm* **571**, 118656 (2019).

20. Jung, L., Hogan, M. E., Sun, Y., Liang, B. M. & Hayward, J. A. Rapid authentication of pharmaceuticals via DNA tagging and field detection. *PLoS One* **14(6**), e0218314 (2019).

21. Bloch, M. S. *et al.* Labeling milk along its production chain with DNA encapsulated in silica. *J Agric Food Chem* **62**, 10615–10620 (2014).

22. Puddu, M., Paunescu, D., Stark, W. J. & Grass, R. N. Magnetically recoverable, thermostable, hydrophobic DNA/silica encapsulates and their application as invisible oil tags. *ACS Nano* **8**, 2677–2685 (2014).

23. Paunescu, D., Stark, W. J. & Grass, R. N. Particles with an identity: Tracking and tracing in commodity products. *Powder Technol* **291**, 344–350 (2016).

24. Pellestor, F. & Paulasova, P. The peptide nucleic acids (PNAs), powerful tools for molecular genetics and cytogenetics. *Eur J Hum Genet* **12**, 694–700 (2004).

## Mathematical verification

### Forbidden combination calculation

Here we verify the equation (1) and (2). Let $h_1, \ldots, h_K$ be the $K$ positions picked by our library. A sequence is not authentic if at these positions we choose nucleotides $N_1, \ldots, N_K$ such that $N_j$ is a nucleotide not allowed at position $h_j$ in the design which has restrictions at $h_j$. There are $3^{K_1} 2^{K_2}$ ways to choose such nucleotides.

Then in total there are $4^{L-K} 3^{K_1} 2^{K_2}$ sequences which are not authentic. Dividing by the total number $4^L$ of sequences, we get

$$\frac{4^{L-K} 3^{K_1} 2^{K_2}}{4^L} = \frac{3^{K_1} 2^{K_2}}{4^K} = \frac{3^{K_1} 2^{K_2}}{4^{K_1+K_2+K_3}} = (1)$$

As for equation (2), note that the probability of a fully randomized sequence to be authentic equals $1 - p$,

and since sequences are independent from each other, the probability that $n$ sequences are authentic equals $(1-p)^n$. To choose $n$, we thus need to solve

$$(1-p)^n = \varepsilon$$

for $n$. This leads to equation (2) by taking the logarithm.

### Nucleotide distribution calculation

Here we verify the equations (3) and (4). To start, note that for each SPOL we produce $l$ sequences, so in total we produce $K\,l$ sequences. Let $q, 1 \leq q \leq L$, be one of the $K$ positions which has been picked. This means there is a design such that at position $q$, only $i$ nucleotides are allowed, with $i$ taking one of the values 1,2,3. By assumption, all $l$ sequences which are produced following this design are produced with the same probability. This, in particular, implies that at position $q$, each of the allowed nucleotides is produced with the probability $1/i$.

Thus, within the $l$ sequences, the expected value of the number of times we see an allowed nucleotide at position $i$ equals $l/i$. Among the other $(K-1)l$ sequences, all four nucleotides appear at position $q$ with probability $1/4$, thus the expected number of occurrences of an allowed letter at position $q$ is $(K-1)l/4$. Summing up, this shows that an allowed letter is expected to occur $(K-1)l/4 + l/i$ many times. To get to the proportion, we still need to divide by the total number $Kl$ of sequences, yielding

$$\frac{(K-1)l/4 + l/i}{Kl} = (3).$$

The proportion equation (4) can be computed analogously, in fact one can deduce equation (4) from equation (3), by using that the proportion of all allowed nucleotides at position $q$ equals $i \times (3)$, thus the proportion of not allowed nucleotides equals $1 - i \times (3)$, and since there are $4 - i$ not allowed nucleotides, we end up with a proportion of $(1 - i \times (3))/(4 - i)$, which is exactly equation (4).

**Total number of sequences per design**

Here we explain how to justify that we may produce more than $(1-p)4^K$ sequences in total. Namely, let U be the total number of DNA sequences. Assume the forger inspects the K picked positions and has access to all U sequences. At the K positions, $4^K$ DNA sequences can be formed.

U should be chosen such that on average at least $p*4^K$ from these $4^K$ sequences are not produced in a fully random CPOL of U sequences. We can think of each of the $4^K$ sequences as a coupon, and then U can be taken to be the expected number of coupons we need to collect (i.e., the number DNA sequences we need to produce) until $(1-p)4^K$ distinct coupons were obtained. This expected number equals $4^K (harm(4^K) - harm(p*4^K))$ which follows from equation (2.6) in ref. 1. Here, $harm(x) = \sum_{i=1}^{x} 1/i$ denotes the harmonic numbers.

In our concrete CPOL, this would allow to produce $9.8083*4^{20} = 1.0784*10^{13}$ sequences in total, an increase by almost factor 10 compared to $(1-p)4^{20} = 1.0995*10^{12}$.

**Reference**

1. Levin, D. and Peres, Y. *Markov Chains and Mixing Times: Second Edition.* (AMS, 2017)